# Combined assessment of auditory distance perception and externalization

*Henning M. Hoppe[1,2], Steven van de Par[2], Virginia Flanagin[3], and Stephan D. Ewert[1,a]*

[1]*Medizinische Physik and Cluster of Excellence Hearing4all, Carl von Ossietzky Universität Oldenburg, 26129 Oldenburg, Germany*

[2]*Akustik and Cluster of Excellence Hearing4all, Carl von Ossietzky Universität Oldenburg, 26129 Oldenburg, Germany*

[3] *German Center for Vertigo and Balance Disorders, University Hospital, Ludwig-Maximilians-Universität München, Germany*

**Abstract**

This study investigates frontal auditory distance perception (ADP) and externalization in virtual audio-visual environments, considering effects of headphone rendering method, room size, reverberation, and visual representation of the room. Either head-related impulse responses from an artificial head or a spherical head model were used for diotic (monophonic) and binaural auralizations with and without real-time head tracking. The visuals were presented through a head-mounted display. Two differently sized rooms as well as an infinitely extending space (echoic and anechoic) were used in which an invisible frontal virtual sound source was located. Additionally, the effect of a freely movable loudspeaker for visually indicating perceived distances was investigated. Both ADP and externalization were significantly affected by room size, but otherwise the two perceptual quantities differed in their outcomes. Room visibility significantly affected ADP, leading to considerable overestimations and more variability in the absence of a visual environment, although externalization was not affected. The movable loudspeaker improved distance estimation significantly, however, did not affect externalization. For reverberation, a (non-significant) trend of improved ADP was observed, however, externalization was significantly improved. Different headphone renderings did not significantly affect ADP or externalization, although a clear trend was observed for externalization.

[a] Email: stephan.ewert_AT_uni-oldenburg.de

# I. INTRODUCTION

Auditory perception provides omnidirectional information about our surroundings, independent of a direct line of sight to a sound source and lighting conditions, contributing to spatial awareness, e.g., guiding head- and eye orientation (Braga et al., 2016). While localization in the horizontal plane is well understood and depends on binaural cues, such as interaural time and level differences (ITDs, ILDs; Middlebrooks and Green, 1991; Wightman and Kistler, 1992, 1997), the picture is less clear for auditory distance perception (ADP; Kolarik et al., 2015; Zahorik et al., 2005), involving several source and environmental factors.

The cues used for ADP are often relative in nature, i.e., they only allow for discrimination between different sound distances (Mershon and King, 1975). One of the most prominent cues for ADP is the sound pressure level. Sound pressure is primarily a monaural cue. In free-field conditions, sound pressure decreases (distance attenuation) by 6 dB per doubling of distance; in reverberant conditions attenuation is lower. However, the original sound source level can change or be unknown, making sound pressure level a relative cue for discrimination of changes in egocentric auditory distance (Zahorik et al., 2005, Mershon and King, 1975).

Studies suggest that distance is also perceived differently, depending on the distance of the sound source. Distances of sound sources in peripersonal space (< 1m) are overestimated, while distances of sound sources in extrapersonal space (> 1m) are underestimated (Kolarik et al., 2015; Zahorik, 2002; Zahorik et al., 2005). Many cues for ADP also differ depending on the distance of the sound source. At source distances above 15 m, spectral cues become important, as higher frequencies are more attenuated by air than lower frequencies (-4 dB / 100 m for 4 kHz, Ingård, 1953, Blauert, 1996), therefore changing the spectral shape of the signal depending on the source distance. The decrease of high-frequency content increases perceived distance when compared to similar sounds with different high-frequency content (Little et al., 1992). Therefore, spectral information is a relative cue for ADP. For sound distances between 1 m and 15 m from the listener, spectral cues do not provide distance information (Kolarik et al., 2015).

Binaural cues can be relevant for distance perception for lateral sources in peripersonal space (< 1m). ITDs are nearly distance independent while ILDs (especially for low-frequencies) can be large at close distances and become nearly independent of distance at farther distances (Brungart, 1999; Duda and Martens, 1998). For lateral sources, the dominant distance cue in peripersonal space is the ILD of low-frequencies (Brungart, 1999) caused by near-field effects in the head related transfer function (HRTF).

In reverberant environments, the direct-to-reverberant energy ratio (DRR; Bronkhorst and Houtgast, 1999; Kopco and Shinn-Cunningham, 2011; Mershon and King, 1975) becomes an additional distance cue. With increasing source distance, the DRR decreases as the direct energy decreases, while the reverberant energy can be approximated by a diffuse sound field and stays roughly constant. DRR thus provides an absolute cue, that allows the judgement of distance based on a single stimulus presentation that can also be used monaurally (Kopco and Shinn-Cunningham, 2011, Mershon and King, 1975).

As a result of these varied and mostly relative cues, estimated distances based on ADP show a stronger variability than visual distance estimated based on visual perception (Anderson and Zahorik, 2014). Because so many different factors are involved in ADP, a systematic congruous evaluation of these cues, especially in reverberant environments and combined with visual cues, requires simulating both the visual and the auditory space in virtual reality (VR).

VR enables the design, simulation and investigation of complex acoustic environments with experimental control, reproducibility and ecological validity (see Keidser et al., 2020, for a review of ecological validity). Headphones allows for precise presentation of various stimuli (e.g., dichotic stimuli), reduces the spatial requirements of the experimental setup and further enables research when a loudspeaker setup is not possible to use (e.g., fMRI studies). The addition of a head-mounted-display (HMD) means audiovisual research questions can be addressed, and it increases immersion. Here we use a VR environment to evaluate the relative contribution of different sound cues for ADP, as well as the influence vision has on ADP.

One major disadvantage of headphone-based stimulus presentation is that the sounds are perceived as internalized, i.e., as originating from a source inside the head (Jeffress and Taylor, 1960; Leclère et al., 2019; Toole, 1970). This contradicts the goal of the simulated sound source to be perceived as originating at a certain distance and direction (externalized) and therefore disrupts realism. Internalization is not caused by the sound presentation over headphones per se (Plenge, 1972, 1974), but depends upon the stimulus properties. For perception to be externalized, the cues, e.g. binaural cues (ITD, ILD, in static positions and during head movements), spectral cues (head-, body-, and pinna filtering effects) and reverberation, must reflect the external sound source. In sound source localization the currently perceived sound signals are compared with stored stimulus patterns (listener expectation). Internalization is likely to occur when incongruencies exist between the listeners expectation and the perceived signals. Externalization is most robust with larger interaural differences, as measured in an anechoic environment. Therefore, sound sources on or close to the median plane are perceived more internalized compared to lateral sources, as the interaural differences are larger (Brimijoin et al., 2013; Leclère et al., 2019).

Reverberation also increases externalization (Begault and Wenzel, 2001; Leclère et al., 2019). A mismatch between the room in which the experiment is conducted (listening room) and the synthesized room in which the signal is presented is referred to as "Room divergence effect", and can cause a significant decrease of externalization (Werner et al., 2016). This decrease of externalization is mainly caused by the auditory mismatch between listening room and synthesized room, rather than the visual mismatch (Gil-Carvajal et al., 2016).

Research on the necessity of individual HRTFs for externalization has led to contradictory results. The importance of reverberation and pinna-cues have been described by Durlach et al., (1992), whereas Loomis et al. (1990) found that approximating the binaural cues to distance and azimuth are sufficient to create the impression of external sound sources, even without the implementation of directional-dependent spectral effects of pinnae. Evidence suggest that artificial HRTFs are sufficient and individual spectral cues are not critical for externalization in more realistic listening situations (Best et al., 2020).

Most real-world listening conditions are dynamic. Dynamic (head) movements of the listener, which cause synchronous changing ear signals, are an important part of auditory perception and lead to externalization, even without Pinnae-related cues. However, only listener induced movements (of themselves or the sound source) cause a benefit (Wightman and Kistler, 1999). If visual and movement dependent context is missing, externalization can only occur if the ear signal corresponds to stimulus patterns in long-term storage (Plenge, 1972).

Head movement only improves the externalization, if the position of the sound source changes relative to the head (Brimijoin et al., 2013). For sound sources with fixed positions relative to the head, externalization decreases. This improvement of externalization by head movements was found to be largest for frontal / back sound source positions (Hendrickx et al., 2017a). This is attributed to the disproportionally larger increase in binaural cues compared to lateral source directions. The achieved externalization from head movements is robust enough to persist even after the movement has stopped. A single ±90° head rotation is sufficient to cause persistent and substantial improvements of perceived externalization in naive listeners (Hendrickx et al., 2017b). For a more thorough review of the topic of externalization see Best et al. (2020).

In the real world, sounds originate from a physical sound source at a certain distance outside of the head and are typically perceived as externalized. Given that ADP cues such as level and DRR are monaural, distances can be judged using diotic presentation (Bidart and Lavandier, 2016; Cubick and Dau, 2016; Prud'homme and Lavandier, 2020). Listeners can judge distances from audio presentations, such as headphones, even if the presentation is not externalized, since ADP cues, such as sound pressure and DRR, can be perceived for non-externalized sounds. We therefore refer to this as inferred distance perception. Performance of inferred distances cannot be used to draw conclusions about the plausibility or authenticity of a simulation and the auralization of an acoustic source within reverberant space. Instead, the source position should be perceived outside the head at a specific location in space (externalized). We therefore asked our participants to judge both distance and externalization in the ADP task.

Vision also affects auditory localization and externalization. The ventriloquist effect is a famous example in which a sound source is erroneously perceived in the direction of a plausible visual stimulus (Jack and Thurlow, 1973). Similar visual capture effects occur in the distance dimension and are stronger for visual stimuli that are closer than the sound source compared to visual stimuli farther than the sound source (Hládek et al., 2013; Zahorik, 2022). Strong visual capture effects have been described in an anechoic environment (Gardner, 1968), however, no visual capture effect was found when reverberation cues were available (Zahorik, 2001). Instead adding vision improved distance judgment accuracy and lowered judgment variability compared to auditory-only stimuli (Zahorik, 2001).

We therefore hypothesize that visual information will decrease the variability of ADP judgments even in a VR environment. Unfortunately, visual distance judgments tend to be underestimated in a VR environment. Kelly et al. (2022) reports distance judgements of 82 % and 75 % of actual distance in VR and 94 % in real world with verbal judgements. Distances above 5 m are underestimated by 38.5 % on average (Korshunova-Fucci et al., 2023). The precision of visual distance perception in VR across a large range of studies has been estimated at 73.48 % of the actual simulated distance (Kelly, 2023). However, for newer head mounted displays (HMDs), distance judgements between 80 % and 86 % are reported.

Another potential effect of the visual environment on ADP results from the central tendency or contraction bias (Hollingworth, 1910; Poulton, 1979) which causes the results of magnitude estimation experiments to be shifted towards the middle of the scale. Comparable effects could occur if the available range of distances in the visual environment, and thus modulated by visual room size, acts as a scale for the listeners.

Taken together ADP and externalization are potentially affected by a several audio-visual cues, with interactions highly relevant for headphone-based virtual acoustic environments and the comparability of perception in such virtual environments with real-life environments. So far, ADP and externalization were not assessed in combination and in a systematic way for different audio-visual environment and parameters of headphone rendering. Headphone-based virtual acoustics has often been applied in motion-restricting environments, such as an MRI scanner, precluding dynamic binaural cues based on head motions. Thus potential effect of dynamic and static headphone rendering are relevant. In this study, we examine the influence of four aspects of an audio-visual simulation on auditory distance perception and externalization, to discover which aspects lead to an authentic simulation in which the distance of a sound source is not only inferred from the auditory cues, but is perceived as being at the simulated distance. The four aspects are: room size, room visibility, headphone rendering technique, and a visually represented loudspeaker for indicating distances. In all conditions an invisible virtual frontal sound source radiated a white noise pulse train in a simulated reverberant room. In a few conditions, the effect of an anechoic environment and signal duration were tested with a reduced number of parameters. Headphone rendering was either diotic (monophonic), binaural using either an extended spherical head model (SHM; Ewert et al., 2021 based on Brown and Duda, 1998) or Head-Related Transfer Functions (HRTFs) of a dummy head. The headphone rendering used six degrees-of-freedom (6-DOF) dynamic motion updates, except for the diotic rendering and a control condition with HRTFs. For the visual environment two differently sized rooms and an infinite unlit space (invisible room) were presented with HMDs and always used 6-DOF motion updates.

Real-time virtual acoustics simulated the reverberation of the two differently sized rooms which were used congruently with their visual representation or for the infinite space. Perceived distances were indicated with a virtual laser pointer on the floor and additionally the role of a visual loudspeaker model for reporting distance was tested.

To account for the visual underestimations in VR, we also performed a visual distance estimation task in which the listeners had to mark specified distances before and after each measurement session. After their final measurement, a questionnaire was given to the listeners which asked about the length of each room, the behavior during the measurement, the helpfulness of the loudspeaker model and the dynamic motion as well as the used strategy to determine the position of the sound source.

## II. METHODS

### A. Listeners

Ten normal-hearing listeners aged between 23 and 32 years (mean: 27.7, SD: 2,8, 6 male, 4 female) took part in the experiment. The threshold for normal hearing was defined as +15 dB HL in the range of 125 Hz to 8 kHz for pure tone detection, and each listener performed an audiometric test starting the first measurement. Five listeners had experience in psychoacoustic experiments.

### B. Audio-visual environments

Two rooms (referred to as big and small) and an infinite space (invisible room) were used as audio-visual environments. The big room was 10 m wide, 16 m long and 4 m high, resulting in a volume of $V_{big} = 640\ m^3$. The small room was 5 m wide, 8 m long and 2.5 m high with a volume of $V_{small} = 100\ m^3$. In the two rooms, the listeners were positioned at half the width of the room and 1 m in front of one short wall, facing into the room. The height of the listeners was either their individual seated height when headtracking was performed or otherwise 1.3 m (representing the average seated height). The visual representation of all environments depicted empty shoebox rooms with neutral grey walls and even lighting. The infinite space was completely unlit and thus black, except for a purple line on the floor. This line indicated the frontal direction and was present in all environments. The length of the line extended to the far wall of the big and small room (16 m and 8 m, respectively) or up to a distance of 60 m in the infinite space. Examples of the visual renderings of the rooms are shown in Fig. 1.

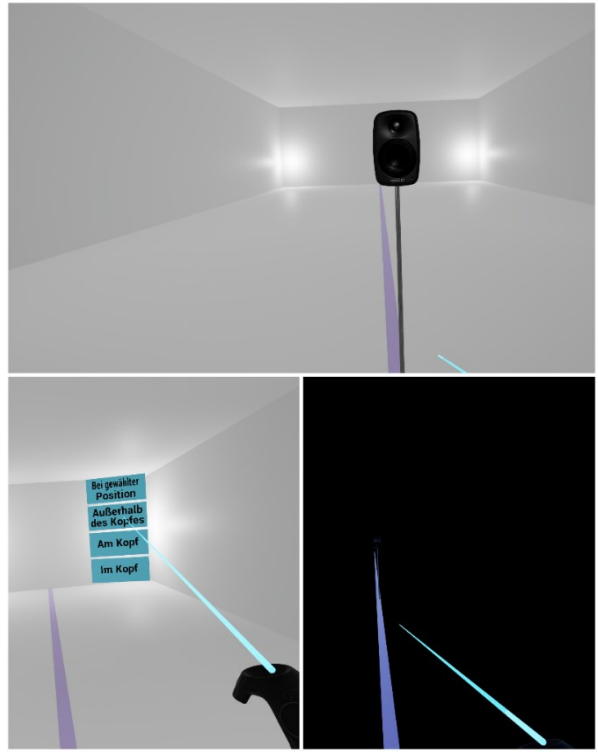

FIG. 1. The upper panel shows the big room with a closely placed loudspeaker for indicating the perceived distance. In the bottom left panel, the controller with laser pointer is shown during the externalization rating in the small room. The bottom left panel shows the listener's view in the invisible room.

The acoustics of the big and small room were simulated using a low-latency, real-time C++ implementation (liveRAZR) of the room acoustic simulator RAZR (Kirsch et al., 2023; Wendt et al., 2014). For early reflections, 3rd order image sources were calculated based on the room geometry and the late reverberation was simulated with a feedback delay network. Auralizations with RAZR have reached a high degree of perceptual plausibility and agreement with real acoustic scenes (Blau et al., 2021; Brinkmann et al., 2019; Wendt et al., 2014). The same room dimensions as for the two visible rooms were used for the room acoustic simulations, resulting congruent room acoustics. The big room had a reverberation time of $T_{60} = 1.2\ s$, the small room $T_{60} = 0.4\ s$. The reverberation time was identical for all octave bands from 250 to 8000 Hz. In the infinite space (invisible room) both room acoustics and an anechoic simulation were used.

For the training phase before the experiment, an independent room was created visually and acoustically. This room was 7 m wide, 12 m long, 3 m high with a volume of $V_{training} = 252\ m^3$ and a reverberation time of $T_{60} = 0.7\ s$.

### C. Apparatus, procedure, and stimuli

During the experiment the listeners were seated in a double-walled acoustic booth wearing an HMD and a hand-held game controller (Valve Index, Valve Corporation, Bellevue, WA), as well as Sennheiser HD650 headphones (Wedemark, Germany). The game controller was used to control a virtual laser pointer (as depicted by the cyan ray in Fig. 1) to indicate the perceived position of the sound source on the floor. The (audio-) visual environment was updated at a rate of 90 Hz based on head tracking. The experiment was conducted in Unreal Engine 4.27 in combination with liveRAZR. The block-size for the audio processing in liveRAZR was 256 samples. To combine the virtual acoustics of liveRAZR with the visuals and motion tracking of Unreal Engine, the "OSC (open sound control)" plugin version 1.1 from Epic Games, Inc. was used to transmit receiver and source positions and rotations to liveRAZR. The stimuli were presented over headphones using an RME FIREFACE UC (Haimhausen, Germany) soundcard. The measurement PC was located outside the booth.

#### *1. Parameters and conditions*

An invisible sound source was placed in the frontal direction (along the purple strip in Fig. 1) at distances from 1 m to 6 m in 0.5 m steps in randomized order in the virtual environments. The listeners indicated the perceived source location on the floor using the laser pointer and by pressing a button on the hand-held controller.

Twenty-eight different conditions were tested in total, varying four main parameters:

*(1) Room size:* The visible and acoustic properties of the room were varied between either the "big" or "small" room.

*(2) Room visibility*. This parameter changed the visual environment in which the listener was situated. "Visible" displayed the room corresponding to the acoustics, while "invisible" displayed the infinite black space.

*(3) Headphone rendering*: Four different renderings methods were used for the sound presented over headphones: (i) "Diotic-static" rendering used only the left ear signals from the dummy head HRTF played on both ears, resulting in a monophonic sound. For the acoustic rendering, no orientation and position (6-DOF) update of the listener's head was used (static rendering for the average seated head position) and only the visuals were updated. (ii) "Binaural-static" rendering used dummy head HRTF for each of the ears and no orientation and position updates. (iii) "SHM-dynamic" rendering used the extended SHM (Ewert et al., 2021) to synthesize the binaural sound with dynamic, real-time 6-DOF orientation and position updates of the listener's head. (iv) "Binaural-dynamic" rendering used a dummy head HRTF (Braren and Kersten, 2022) to synthesize a binaural sound with dynamic, real-time 6-DOF updates. For performance reasons, the resolution of the HRTF had to be reduced to 890 directions in total with a frontal resolution of 5 degrees (see Appendix A).

*(4) Loudspeaker model:* This parameter enabled or disabled the display of a visible loudspeaker model on a pole with a height of 1.3m (referred to as LS in the following). The loudspeaker model was randomly positioned within the possible range of source positions at the start of the stimulus presentation. During and at the end of the presentation, its position on the floor was controlled by the virtual laser pointer to visually represent the perceived position.

Given the high number of parameters, not all combinations could be measured in a full factorial manner. Table I depicts the four sets of conditions for which a subset of parameters was tested in a fully factorial way. In addition to the upper parameters, an anechoic condition was tested for the invisible room in set C. Moreover, the signal duration was varied as a parameter in set D.

TABLE I. Fully factorial condition sets using a subset of the four main parameters. Varied parameters are highlighted in italics. Additional parameters are provided in the lower row for sets C and D.

|  | Set A | Set B | Set C | Set D |
|---|---|---|---|---|
| Room size | *Big and small* | Big | Big | Big |
| Room visibility | *Visible and invisible* | *Visible and invisible* | Invisible | Visible |
| Headphone rendering | *All* | *All* | Binaural - dynamic | Binaural - static |
| Loudspeaker model | With | *With and without* | *With and without* | *With and without* |
| Additional |  |  | *Reverberant and anechoic* | *2 s and 6 s signal duration* |

Before the start of a new condition, a large floating sign displayed "static" or "dynamic" to let the listeners know when the dynamic head-tracking of the acoustics was available. To let the listeners get used to the different scenes, the conditions were not fully randomized but rather presented in pseudo randomized order according to room size, room visibility, and display of the loudspeaker model e.g. big visible room with LS followed by small invisible room without LS. In each environment all rendering techniques were tested in pseudo randomized order before a change of environment occurred.

## 2. Measurements

The experiment was conducted in three sessions, with each session lasting around 2 hours. In each session every condition was measured and every source position was tested twice. Source positions were randomly presented at a distance between 1 m and 6 m (step size 0.5 m). This led to 616 measurements per session and 1848 total distance estimations and externalization ratings per listener.

To mark the perceived distance, the listeners had to point the virtual laser on the floor and press the trigger-button. To confirm their input, after another button press a virtual confirm button was spawned that had to be clicked via pointing of the laser pointer and pressing the trigger button. In conditions with the LS model (see upper panel of Fig. 1), its position was changed in response to the listener's perceived source location as indicated with the laser pointer. This was possible during and after the stimulus presentation. In conditions without LS, no visual feedback about the marked position was provided until after the stimulus presentation. Then a small object on the floor, similar to a black hockey puck, indicated the marked position. After the distance estimation, the listeners were asked to rate the perceived externalization of the stimulus. For this, four buttons labeled "In the head", "At the head", "Intermediate", "At the position" (original German labels: "Im Kopf", "Am Kopf", "Außerhalb des Kopfes", "Bei gewählter Position") were shown on a virtual display board (see lower left panel in Fig. 1) and the listener had to select one rating via pointing of the laser pointer and confirmation with the trigger button.

Prior to the measurements, eight stimuli were presented over headphones without the HMD to familiarize the listeners with the sound level range and give them an impression of externalization. The stimuli presented were the nearest and farthest source position for the big and small room, first presented binaurally, then diotically. Each measurement session started with a training phase of the two full conditions [visible room, Binaural dynamic, with / without loudspeaker], tested in the independent training room.

Before the first condition and after the last condition of a measurement session was finished, a visual distance indication task was conducted. Participants indicated the distance of 5 m in the small room, 5 m and 10 m in the big room and 5 m, 10 m and 15 m in the invisible room. Responses were given by pointing the laser pointer at the intended position and then pressing the trigger-button.

After the last measurement session, a questionnaire was given to the listeners in which they had to estimate the length of the visible rooms and the length of the purple line in the invisible room. Furthermore, they rated if they had their eyes open during the measurement, their usage of the dynamic head movements, and the helpfulness of the LS model for indicating distances. In the last part of the questionnaire the listeners were asked about their general strategy to judge the sound source distance. For the estimation of the room sizes and their distance judgement strategies the questionnaire contained empty fields. All other responses used a numerical rating scale from 1 to 5 (only whole numbers) where 1 corresponds to a positive and 5 to a negative rating.

### 3. Stimuli

The stimulus radiated by the virtual sound source was an alternating series of frozen white noise bursts (20 ms) and pauses (300 ms) with 155 ms pause in the beginning and at the end. This led to a series with 3 bursts per second. For dynamic conditions (SHM and binaural-dynamic) the overall stimulus duration was 6 s. For the static renderings, where listener's movements had no effect, the stimulus duration was shortened to 2 s. The stimulus was presented to the listeners using a simulated (invisible) directional sound source rendered in real time. The different rendering methods were realized with liveRAZR (see Sec. II. B.). For the HRTF-based renderings (diotic-static, binaural-static and binaural dynamic) a headphone equalization was done, using the inverse frequency response of the headphones. The stimulus was calibrated in the big room for the diotic, SHM and binaural rendering at the farthest source distance to a level of 61 dB SPL. This resulted in levels of 72 dB SPL for diotic and binaural rendering and 75 dB SPL for the SHM rendering at the closest distance. The anechoic condition had levels of 56 dB SPL for the farthest and 72 dB SPL for the closest distance. With the same sound source parameters, the levels in the small room were 62.5 dB SPL for the diotic and binaural rendering at the farthest distance and 72 dB SPL at the closest. For the SHM rendering the levels were 64 dB SPL at the farthest and 75 dB SPL at the closest distance.

### D. Statistical analysis

Data were analyzed using a 3-way repeated measures analysis of variance (rmANOVA) for set A and B and a 2-way rmANOVA for set C and D in IBM SPSS statistics (Version: 29.0.0.0 (241)). Before the analysis, the data was searched for outliers exceeding 2 standard deviations from the mean. The individual listener results were searched for each source distance in each condition. From the 18480 measured positions 52 (0.28 %) were detected as outliers and removed from the data. For individual listeners the maximum number of removed measured positions for one source distance within one condition is 1, the maximum per condition is 3. Afterwards, the remaining data of each listener per source distance and condition were averaged and taken as final result. These were searched for outliers between the listeners. Here one listener was excluded from the analysis of ADP. For distance perception, a second order (quadratic) Bézier curve with two parameters, slope, and bending (see Appendix B), was fitted to the results and these parameters for each condition were subjected to the statistical analysis. For a bending parameter of 0, the slope equals that of the linear regression. Thus slopes larger than 1 indicate an overestimation in perceived distance, slopes smaller than 1 indicate an underestimation of distance. A bending larger than 0 indicates that the slope is steeper for close distances than for farther distances (decreasing slope with distance), and vice versa for negative bending. The analysis of the externalization was performed on the combined number of "In the head" and "At the head" ratings averaged over all source positions per condition.

## III. RESULTS

### A. Auditory distance perception

In general, listeners were able to judge relative distances, i.e., for all rooms and headphone renderings, larger distances were perceived larger than smaller distances. The average perceived distances across all listeners are shown in Fig. 2, with error bars at one standard deviation. For readability, the standard deviation is only indicated in one direction and the graphs are slightly shifted along the x-axis. Data are separated by room size, the presence of a loudspeaker (columns) and room visibility, as well as the additional anechoic and signal duration conditions (rows). Different rendering techniques are indicated by color. Yellow represents diotic-static rendering, green the SHM-dynamic rendering and red and blue the binaural-dynamic and binaural-static rendering, respectively.

The distance of the perceived sound source was generally overestimated in all conditions, except for the visible small room (Fig. 2c). This can also be seen in Table II where the slope and bending parameters are listed using the same layout as Fig. 2. The slopes were almost all >1 corresponding to an overestimation. The overestimation ranges from about 1.6 times the actual sound source distance, e.g., 6 m in the big, visible room with LS and binaural-dynamic rendering is estimated at 10.1 m (Fig 2b, red line), to above 3 times the actual distance, e.g., 1 m in the big, invisible room without LS and with binaural-dynamic rendering is estimated at 3.2 m (Fig. 2d, red). The distances of closer sound sources are more greatly overestimated than the distances of farther sound sources. For instance, 1 m source distance in the big visible room without LS and with binaural-dynamic rendering is estimated at 3.12 m (overestimation > 3), whereas the 6 m source distance was estimated at 11.29 m (overestimation < 2, Fig. 2a, red line).

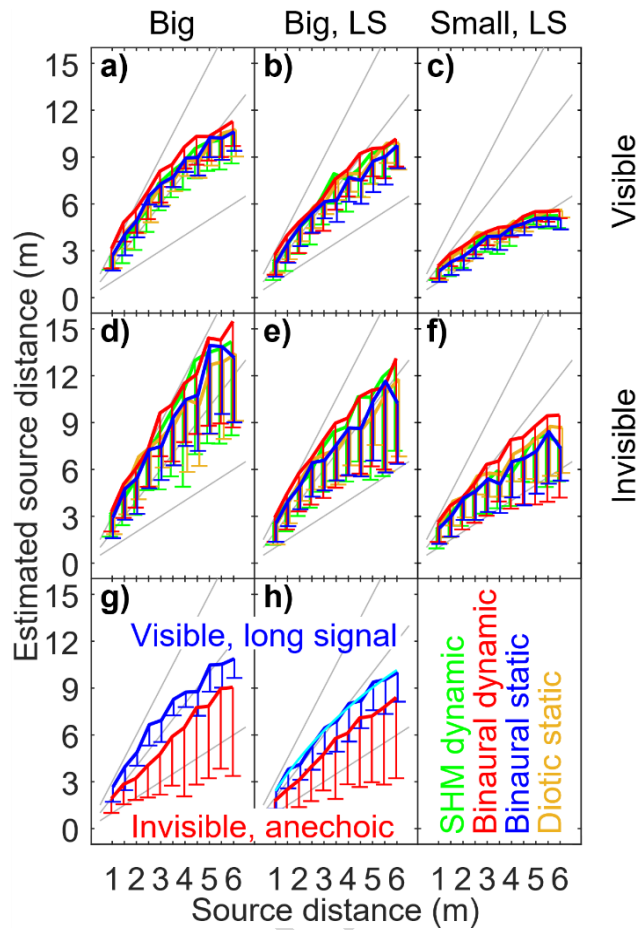

FIG 2. Estimated vs actual source distances. Mean estimated distances are plotted with error bars at one standard deviation for all tested parameter combinations. The columns indicate the simulated room size and the availability of the freely movable loudspeaker model (LS). The upper and middle row indicate the visibility of the room. The bottom row shows the results of the anechoic (panel g and h, red line) and long duration (panel g and h, blue line) stimuli. The rendering technique is indicated by color. The grey straight lines indicate slopes of 1, 2, and 3. For better readability, the standard deviation is only shown in negative direction and a slight shift on the x-axis was introduced. The cyan curve in panel h) shows an example of the Bézier curve fitted to the averaged distance estimations, as used for the statistical analysis.

Perceived distances were larger and more overestimated in the larger room. In the 3-way rmANOVA [Set A: Room size (big / small) x Rendering (diotic-static / binaural-static / SHM-dynamic / binaural-dynamic) x Room visibility (visible / invisible)] there was a significant main effect of room size in both the slope and bending parameters [$F(1,8) = 44.877$, $p < 0.001$ & $F(1,8) = 21.508$, $p < 0.002$ respectively]. The small room had a mean slope of $1.01 \pm 0.34$ and the big room had a mean slope of $1.69 \pm 0.61$. Distance perception was also significantly more precise in the small room. The bending values were larger in the small room $0.26 \pm 0.10$ than in the big room $0.19 \pm 0.12$. The higher positive bending for the small room indicates that farther sound sources are perceived closer in the small room compared to the big room. No main effect of rendering or visibility and no significant interaction for slope or bending were found.

For room visibility (compare Fig. 2a-c to d-f) an increase in perceived distance can be observed for the invisible room. This is especially noticeable for medium to far sound source position. For the big visible room, the distances for sound sources between 3 to 6 m are overestimated by around factor 2.8 to factor 2 (without LS) and around factor 2 to 1.6 (with LS) while in the invisible room the overestimations range from around factor 3 to 2.4 (without LS) and around factor 2.4 to factor 2 (with LS). In the small visible room, only the distances of closer sound sources are overestimated (around factor 2) while the farther source distances are perceived quite correct (close to factor 1), e.g. 6 m were perceived as 5.58 m and 5.5 m as 5.5m with binaural-dynamic rendering (Fig. 2c, red line). In the invisible small room, source positions are overestimated by a factor of around 2.5 for close source positions and with increasing source distance the overestimation decreases to a factor of around 1.5 for the farthest sound source. Also noticeable is the difference in consistency of the estimated distances. In the visible rooms the estimated distances across listeners are relatively constant for all virtual sound source positions. In the invisible room however, the estimated distances vary strongly between and within the listeners and the variation increases with increasing source distance. The 3-way rmANOVA [Set B: LS (with / without) x Rendering (diotic-static / binaural-static / SHM-dynamic / binaural-dynamic) x Room visibility (visible / invisible)] indicated a significant main effect of the LS and a significant interaction between the LS and room visibility for slope. For bending no significant main effect or interaction was found. With a mean slope of 1.99 ± 0.62 (without LS) and 1.69 ± 0.61 (with LS), the distance estimation is significantly improved by the presence of the loudspeaker [$F(1,8) = 6.348$, $p = 0.036$] The interaction between the factors LS and room visibility is significant with $F(1,8) = 6.538$, $p = 0.034$. A post-hoc analysis revealed a significant improvement of distance estimation for LS in the invisible room [mean slope: 2.32 ± 1.00 (without LS), 1.87 ± 1.00 (with LS), $p = 0.033$]. Additionally, a significant improvement of distance estimation was found for the visible room compared to the invisible room without LS, with mean slopes of 1.67 ± 0.30 (visible room) and 2.32 ± 1.00 (invisible room), $p = 0.039$.

Interestingly, no significant differences in perceived distance or the precision of perceived distance were found between rendering techniques (colors in Fig. 2a-f). The estimations with binaural-static rendering are a little more veridical compared to the binaural dynamic rendering. However, the differences between the estimated distances for the rendering techniques lie within one standard deviations of each other. Especially in the invisible room, where the differences are most prominent, the standard deviations are also the largest.

The presence of the LS model (compare Fig. 2a, d to b, e and g to h) to indicate the acoustically perceived distance reduced the overestimation in both the visible and invisible rooms, with a stronger effect in the invisible room.

As expected, the perceived distances clearly differed between reverberant and anechoic conditions (Fig. 2d, e and g, h, red lines). Sources in an anechoic space are perceived considerably closer than the reverberant ones, while distances still being overestimated. Without the LS, the estimated distance is roughly halved if the stimulus is presented anechoically compared to the reverberant version, e.g. in the big, invisible room with binaural dynamic rendering without LS the source distances: 2 m, 3.5 m 6 m are marked at 6 m, 10.1 m 15.5 m with reverberation (Fig. 2d, red line) and at 3.2 m, 5.9 m and 9 m when presented anechoically (Fig. 2g, red line). The presence of the LS partially alleviates the difference between anechoic and reverberant rooms, but the difference is still clearly observable. A 2-way rmANOVA [Set C: LS (with / without) x reverberation (reverberant / anechoic)] indicated a significant main effect of the LS model for the slope. With a mean slope of $2.01 \pm 0.98$ without LS and $1.64 \pm 0.95$ with the LS, the distance estimation was significantly improved by the LS ($F(1,8) = 12.962$, $p = 0.007$). No further significant main effects or interactions were found for slope and bending.

Regarding the signal duration (Fig. 2a, b and g, h, blue lines), only minor differences in perceived distance for the 2 s and 6 s signal duration were observed, e.g. source distances of 3 m and 6 m in the big, visible room with diotic-static rendering and without LS were marked at 8.3 m and 10.8 m with the 6 s signal (Fig. 2g, blue line) and at 7.75 m and 10.6 m with the 2 s signal (Fig. 2a, blue line). A 2-way rmANOVA [Set D: [LS (with / without) x Signal duration (2 s / 6 s)] showed no significant main effects or interactions for slope or bending.

TABLE II. Average fitted slopes and bending (in parentheses) parameters of the rendering techniques in the different conditions (same layout and colors as in Fig. 2. Sd: SHM dynamic, Bd: Binaural dynamic, Bs: Binaural static, Ds: Diotic static.

| Big | | | | Big, LS | | | | Small, LS | | | | |
|---|---|---|---|---|---|---|---|---|---|---|---|---|
| Sd | Bd | Bs | Ds | Sd | Bd | Bs | Ds | Sd | Bd | Bs | Ds | |
| 1.75 ±0.35 (0.24 ±0.08) | 1.69 ±0.37 (0.33 ±0.19) | 1.62 ±0.28 (0.26 ±0.10) | 1.62 ±0.34 (0.23 ±0.12) | 1.57 ±0.29 (0.19 ±0.12) | 1.54 ±0.33 (0.22 ±0.14) | 1.45 ±0.33 (0.15 ±0.19) | 1.48 ±0.38 (0.21 ±0.12) | 0.73 ±0.09 (0.30 ±0.20) | 0.75 ±0.12 (0.32 ±0.16) | 0.72 ±0.18 (0.18 ±0.18) | 0.75 ±0.15 (0.34 ±0.10) | Visible |
| 2.46 ±1.28 (0.14 ±0.10) | 2.48 ±1.20 (0.18 ±0.19) | 2.25 ±0.87 (0.13 ±0.09) | 2.07 ±0.81 (0.11 ±0.30) | 2.07 ±1.16 (0.12 ±0.23) | 1.93 ±1.19 (0.23 ±0.19) | 1.71 ±0.86 (0.23 ±0.16) | 1.77 ±0.94 (0.19 ±0.30) | 1.23 ±0.48 (0.20 ±0.24) | 1.47 ±1.03 (0.21 ±0.19) | 1.16 ±0.54 (0.23 ±0.18) | 1.26 ±0.58 (0.27 ±0.27) | Invisible |
| - | 1.53 ±1.11 (0.06 ±0.18) | 1.69 ±0.33 (0.25 ±0.10) | - | - | 1.35 ±0.98 (0.12 ±0.07) | 1.57 ±0.47 (0.17 ±0.14) | - | - | - | - | - | Invisible-Anechoic Visible-Long sig. |

Taken together, auditory distance perception was significantly affected by room size. The presence of a visible LS model improved distance estimation, more so in the invisible room than in the visible room, suggesting that vision does have a positive effect on ADP. Rendering technique did not show any significant effects on ADP.

**B. Externalization**

Externalization was influenced by room size, headphone rendering technique and to a small degree the presence of the LS. In Fig. 3 the averaged externalization ratings across listeners and all source positions are shown in the same layout as for ADP in Fig. 2. The underlying individual data for the 10 listeners are based on 6 ratings for each of the 11 distances (66 ratings per listener in total) per condition. The black error bars indicate inter-individual standard deviations within each externalization category (for better readability, only the lower standard deviation is shown in the top bars). The white horizontal lines and white error bars show the average and standard deviation of the pooled internalized "In the head" and "At the head" ratings (indicated in red color tones), separating the externalized "Intermediate" and "At the source" ratings (indicated in blue color tones).

Overall, listeners externalized the sounds more in the big room (with LS) compared to the small room (Fig. 3, panels b, e and c, f). The rendering techniques show clear effects on the externalization. The combined rating of "In the head" and "At the head" (Fig. 3) is considerably higher for the diotic-static rendering, also the "In the head" ratings alone are the highest for the diotic-static rendering compared to the other rendering techniques. The results also show a better externalization of dynamic renderings compared to static ones. Binaural-dynamic rendering was externalized most often (5.68 % ± 3.02 % internal perception), followed by the dynamic SHM (13.21 % ± 4.94 % internalization) and the binaural-static rendering (17.10 % ± 7.30 % internal perception). The worst externalization was perceived with the diotic-static rendering, which had an internal perception rating of 32.45 % ± 4.27 %. A 3-way rmANOVA [Set A: Room size (big / small) x Rendering (diotic-static / binaural-static / SHM-dynamic /

binaural-dynamic) x Room visibility (visible / invisible)] indicated a significant main effect of room size. With mean internalization ratings of 0.79 ± 0.74 (13.17 % ± 12.33 %) (big room) and 1.33 ± 1.06 (22,17 % ± 17.67 %) (small room), stimuli in the big room were significantly more externalized [F(1,9) = 14.875, p = 0.004].

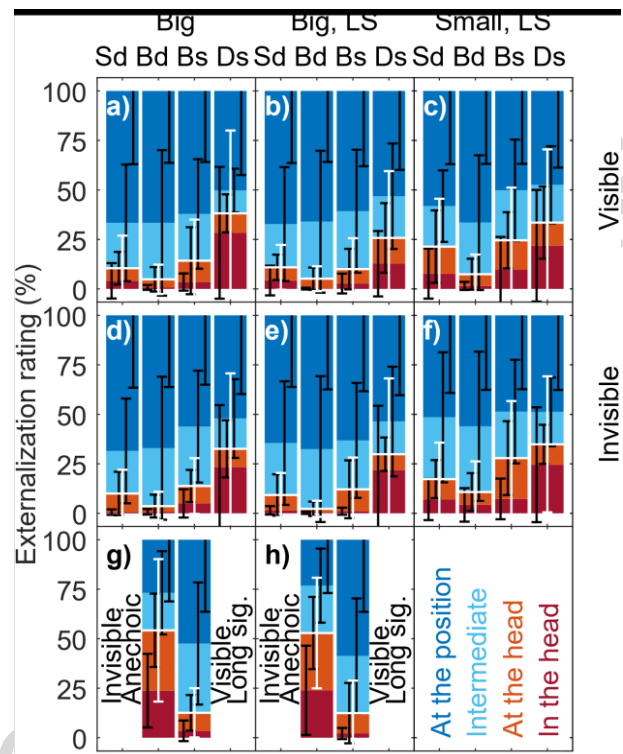

FIG. 3. Averaged externalization ratings over all listeners and virtual source distances for the different conditions (same panel layout as in Fig. 2) The used rendering techniques are shown above the bars (Sd: SHM dynamic, Bd: Binaural dynamic, Bs: Binaural static, Ds: Diotic static) and the colors indicate the rated externalization. The black error bars show the standard deviation of each rating category and the white error bars show the standard deviation of the pooled "In the head" and "At the head" ratings, which were used to evaluate the externalization.

Between the visible and invisible rooms only small differences in externalization can be observed (Fig. 3, a - c and d - f). Small differences in externalization are also observable between the presence and absence of the LS in all conditions, except the diotic-static rendering in the visible room (Fig. 3 a and b, column "Ds"). Here the perceived externalization increased from 61.82 % without LS to 74.24 % with the LS. The 3-way rmANOVA [Set B: LS (with / without) x Rendering (diotic-static / binaural-static / SHM-dynamic / binaural-dynamic) x Room visibility (visible / invisible)] revealed no significant main effects or interactions. Listeners externalized the sound considerably less in anechoic conditions than in the corresponding reverberant conditions (53.56 % ± 0.96 % internal perception in the anechoic condition (Fig. 3g, h) and 2.95 % ± 0.96 % internal perception with reverberation (Fig. 3d, e)). In the 2-way rmANOVA [Set C: LS (with / without) x reverberation (reverberant / anechoic)] was a significant main effect of reverberation. With mean internalization ratings of 3.21 ± 1.88 (53.5 % ± 31.33 %) in the anechoic conditions and 0.177 ± 0.33 (2.95 % ± 5.5 %) in the corresponding reverberant conditions, the externalization in anechoic conditions was significantly less [$F(1,9) = 28.876$, $p < 0.001$].

The duration of the stimuli did not affect externalization. The 2 s stimuli and the 6 s stimuli in the binaural-static control condition were similarly externalized, with 12.12 % ± 3.00 % (2 s) and 12.50 % ± 0.11 % (6 s) internal perception. The 2-way rmANOVA [Set D: LS *(*with / without) x Signal duration (2 s / 6 s)] showed no significant main effects or interactions.

To further assess the effect of source distance, additionally, the externalization ratings of all listeners for the different source distances are shown in Fig. 4, again using the same panel layout as in Figs. 2 and 3. The percentage scale is based on the 60 ratings, resulting from 6 ratings from 10 listeners per condition and source distance. Here, a dependency of externalization on source distance can be observed. Closer sound sources were perceived less externalized than farther sound sources. This increase of externalization with distance is more pronounced for the closer source distances and flattens at around 3 m distance. No statistical analysis was performed here.

In summary, although the headphone rendering technique, in particular the ability to perform head movements, visibly changed the amount to which listeners externalized the sound, the differences were not significant. Only the presence of reverberation and larger room size significantly increased externalization.

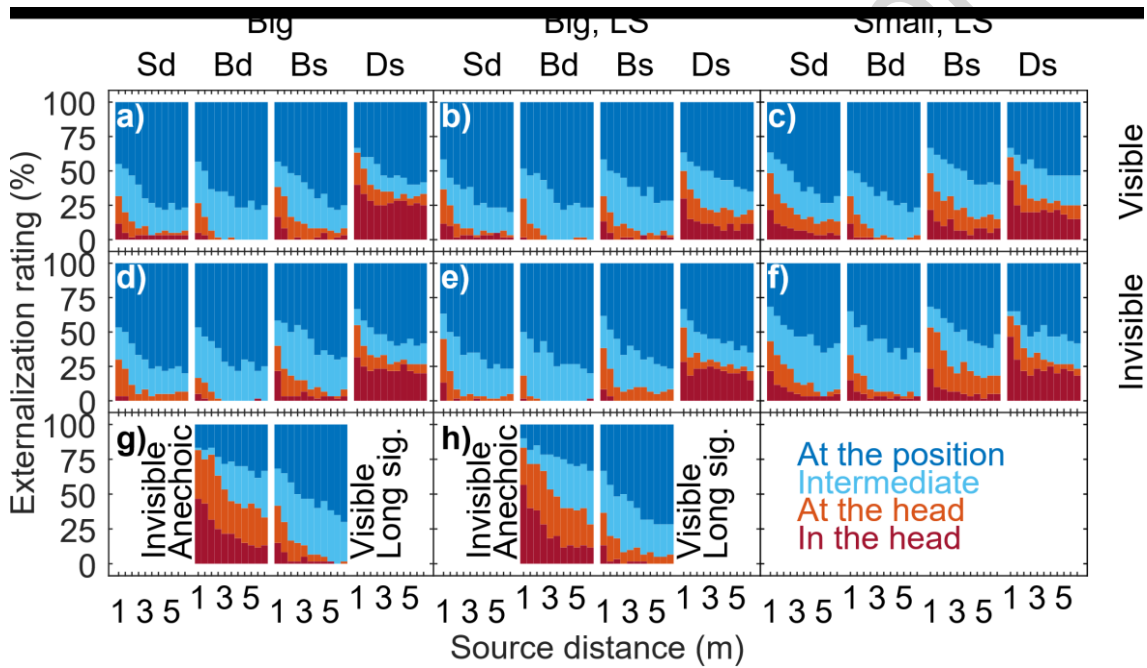

FIG.4. Externalization ratings of all listeners as a function of source distance in the same layout is as in Fig. 2, and color scheme as in Fig. 3. The percentage scale is based on 60 externalization ratings for each distance. Rendering techniques are plotted in separate columns with the following two-letter labels: Sd: SHM dynamic, Bd: Binaural dynamic, Bs: Binaural static, Ds: Diotic static.

## C. Visual distance perception

To assess the influence of the visual virtual space on the results, listeners performed a visual distance estimation task and in a questionnaire were asked how long each room was.

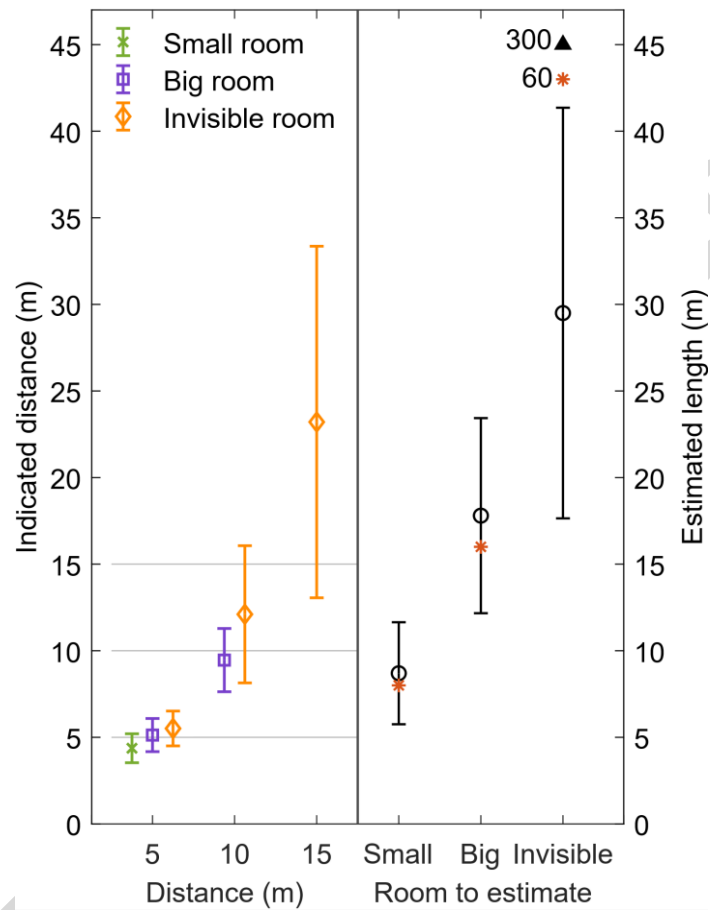

FIG. 5. Mean estimates of visual distance indication task (left) and mean estimated room sizes (right). In the left panel, the "x", "square" and "diamond" symbols indicate the averaged results of the visual distance indication task with one standard deviation. In the right panel, the circles show the averaged estimated room sizes with one standard deviation and the asterisks indicate the real room size. One rating of a room size of 300 m for the invisible room is considered an outlier (indicated at the top of the right panel) and therefore not included in the average.

The "x", "square" and "diamond" symbols in the left panel of Fig. 5 show the averaged estimations of the visual distance estimation task with error bars (one standard deviation). The 5 m distances were judged relatively precisely in all rooms, whereas the larger the target distance became, the larger the deviation and variability.

The marked distance for 5 m in the small room (green x) was 4.37 m ± 0.84 m (87,5 % / -12.5 %). In the big room (purple square) the marked distances were 5.13 m ± 0.96 m for 5 m (102,6 % / +2.6 %) and 9.46 m ± 1.83 m for 10 m (94,6 % / -5.4 %). In the invisible room (orange diamond), the 5 m distance was marked at 5.51 m ± 1 m (110,2 % / +10.2 %), the 10 m distance at 12.1 m ± 3.96 m (121 % / +21 %) and the 15 m distance at 23.21 m ± 10.15 m (154,3 % / +54.3 %).

Similarly, when participants were asked how long each room was in a post-experiment questionnaire (see black circles in the right panel of Fig. 5), the deviation also becomes larger the larger the room (increasing from left to right). The participants estimated a room length of 8.7 m ± 2.95 m (108,7 % / +8.7 %) for the small room. For the big room a length of 17.8 m ± 5.63 m (111,26 % / +11.26 %) was estimated and the length of the purple line in the invisible room was estimated to be 29.5 m ± 11.86 m (49.17 % / -50.83 %). The invisible room estimation is based on 8 ratings only. One participant did not give a rating and the 300 m estimation is considered an outlier (300 > mean + 2 standard deviations). The red asterisks indicate the real length of the rooms (or purple strip for the invisible room).

**D. Self-reported distance judgement strategy and cue assessment**

Individual ratings from the numeric 5-point questionnaire are shown in Fig. 6. The evaluation of the rated questions contains only 9 participants, since the questionnaire was changed after the first participant only had "yes / no" response alternatives and always answered "yes". All participants confirmed that they had their eyes open during the experiment (Median = 1: yes). Although listeners' median rating was 3 for using the dynamic motion cues, they rated their usefulness quite high (median = 2). Listeners rated the helpfulness of the visible loudspeaker high, with a median rating of 2. More listeners rated helpfulness of the LS higher than that of the dynamic motion. Only one participant found it less helpful than the dynamic motion.

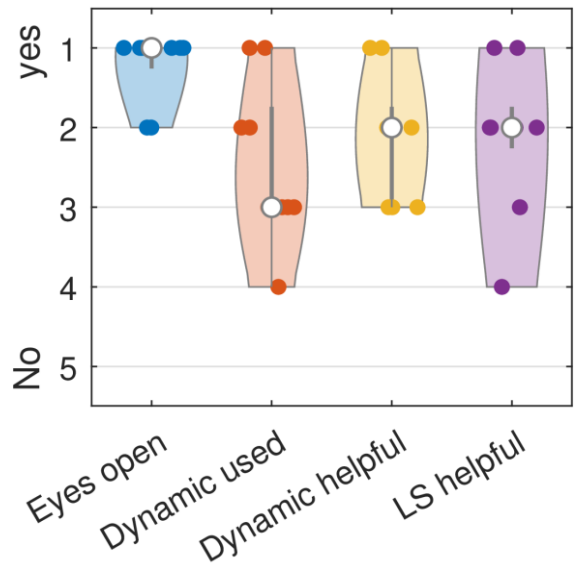

FIG. 6. Results of the questionnaire on a numeric 5-point scale with 1 corresponding to a positive and 5 to a negative rating. The questions are indicated on the x-axis, covering whether the listeners had their eyes open during the measurement (eyes open), whether they used the dynamic movement (dynamic used), and how helpful they found the dynamic movement (dynamic helpful) and the loudspeaker model (LS helpful). Colored dots show individual listener ratings (with some data points overlapping). The white dots show the median rating.

The specific strategies listeners used to determine the distance to sound sources are listed in Table III. Multiple answers per listener were possible. The most frequently mentioned strategy, which was stated 4 times, was "intuition". The next most common strategy, "listening to reverberation", was mentioned 3 times. After that, with 2 mentions each, came "loudness comparison", "comparing consecutive stimuli", and the "comparison of the perceived sound to an imaginary loudspeaker" (in case of conditions without LS) or the usage of the laser pointer on the floor as visual comparison to the auditorily perceived distance. The least mentioned strategy was sorting the perceived stimulus in the categories "close", "medium", and "far". This strategy was adopted by one listener.

Table III. Used strategies and their frequency according to the 10 listeners. Multiple answers were possible.

| Strategy | Frequency |
|---|---|
| Intuition | 4 |
| Use reverberation | 3 |
| Compare loudness | 2 |
| Compare consecutive stimulus presentations | 2 |
| Slide an imaginary loudspeaker along the line and compare the imaginary image with the perceived sound | 2 |
| Make categories: close, mid, far | 1 |

## IV. DISCUSSION

### A. Auditory distance perception

In the ADP task, distance of the sound source was overestimated by all listeners. In the big room (10 x 16 $m^2$, 640 $m^3$) overestimation was the greatest and present over the entire range of source distances. In the small room (5 x 8 $m^2$, 100 $m^3$), visibility played an important role, leading to more precise estimations of the farther source distances. The increased reverberation in the big room and resulting lower DRR could have caused an increase in perceived auditory distance compared to the small room.

The previously described underestimation of distances for sound sources farther than 1 m (Kolarik et al., 2015; Zahorik et al., 2005) could not be observed. A similar overestimation of auditory perceived distance as in the current data is reported by (Kroczek et al., 2023). A systematic underestimation of visual distances in VR (Kelly et al., 2022; Kelly, 2023;

Korshunova-Fucci et al., 2023) cannot explain the here observed overestimation of source distances, as the participants were able to indicate the visual distance relatively accurately in the range of acoustically presented distances.

Another reason for the observed increase in estimated source distance with increased visual room size could be the central tendency (Hollingworth, 1910) and scale effects in general (Poulton, 1979). If the listeners used the full length of the room as visual reference frame or distance scale, the center tendency would result in closer and farther positions being shifted towards the center of the scale. Therefore, the overestimation of the closer sound source distances would be explained by the tendency to place results towards the middle of the scale and the overestimation of the farther sound sources in the large room and the infinite space would be explained by using an incorrect (too long) distance scale.

The presence of a visual loudspeaker during the distance perception experiment improved performance, especially for medium and farther source distances. The loudspeaker provided a possibility for an auditorily perceived distance to be compared to a visual representation of the sound source. A "visual capture effect" might also be relevant. This effect is asymmetrical, i.e., sounds behind a visual object get "captured" more strongly than sounds in front (Zahorik, 2022). It is most prominent at farther sound sources and thereby explain the improvements especially in the medium and farther ranges of simulated source positions. Given that the position of the loudspeaker was interactively adjusted by the participant to match the perceived distance, the role of a potential capture effect remains unclear.

Despite the substantial differences in the rendering techniques used, no significant difference in ADP between rendering techniques were found. A frontal sound source, such as the one used in this study, does not cause binaural cues in the direct sound. The symmetrical positioning of the listener and sound source within the rooms also results in nearly identical early reflections at both ears. As sound from distances larger than 1 m do not lead to distance-relevant binaural cues (Brungart et al., 1999), the dynamic rendering in the current study likely provide few to none additional cues for ADP. As spectral cues have no effect on ADP in the range of 1 m to 15 m (Kolarik et al., 2015), these cues were also not relevant. The remaining relevant cues for ADP were intensity and DRR, which both can be considered monaural cues, and are thus well audible even in the simplest rendering method used here. This is likely why similar performance was observed for all rendering techniques. While individualized HRTFs were not considered in the current study, strong differences in performance with individual HRTFs appear questionable for the above mentioned reasons.

In the anechoic conditions, distance could only be judged using intensity cues. Nonetheless perceived distance increased with simulated distance, however the high direct energy resulted in closer perceived source distances compared to their reverberant equivalents, while still being overestimated.

**B. Externalization**

The externalization ratings were generally high in all conditions. Significant differences occurred between the big and small room and reverberation or anechoic acoustics. One reason for the generally high externalization ratings might be the simultaneous presentation of the stimulus in a acoustically and visually congruent simulated rooms. This likely created a plausible scenario and avoided the room divergence effect, therefore supporting externalization (Li et al., 2021). Concerning the negative effects of room divergence on externalization (Werner et al., 2016), the current results indicate that the missing visual information in the invisible room did not cause a mismatch. A possible explanation might be that the listeners know the (visible) rooms from other conditions. This way the presented stimuli finds a matching comparison (when comparing perceived cues to expected cues) even in the invisible room.

It was unexpected that even the diotic-static rendered stimuli received an externalization rating of around 67 %, given that diotically presented stimuli are typically perceived internal. One reason for this high externalization ratings might be the frontal source direction, generally providing relatively limited binaural cues. Accordingly, diotically rendered sounds did not differ much from the binaural ones, if listeners were not moving or rotating their head. With head movements, the missing binaural cues should have been perceivable to the listeners and externalization should have been affected. Thus, the observed high externalization rating of the diotic stimuli likely resulted from a mostly static listener position. Another reason for the high externalization ratings might be related to the randomly interleaved presentation with dynamic conditions, while the room stayed the same. Once an external perception was achieved, it was likely upheld between the different source distances and even between rendering conditions, despite the absence of binaural cues. Moreover, supported by a reduction of head movements engaged during performing the experiment. However, head motion had not been tracked during the experiment, and thus reduced head motions stay so far in informal observation by the experimenter, supported by self-reports of some listeners.

The effect of the presence or absence of reverberation on externalization is clearly visible in the results (Figs. 3, 4, a, b vs. g, h). As completely anechoic sounds are unusual in daily-life, they sound unnatural and violate the listeners' expectation (perceived signal cues do not match any known cue combination). Additionally, without reverberation no absolute spatial cue is present, which can cause the anechoic stimuli to be perceived at no distance (resulting in "in the head" perception). However, some listeners were able to externalize the anechoic sounds. Here, most likely the dynamic motion and the resulting binaural cues caused the externalization (Hendrickx et al., 2017a; Plenge, 1972).

Regarding the use of individual HRTFs, an improved externalization in dynamic and reverberant simulations is questionable. In comparison to the SHM, the addition of pinna and torso-related cues in the current dummy head HRTF resulted in only small improvements. Individual pinna and torso cues might have some effect. However, it was shown that dynamic binaural cues caused by motion, are sufficient to cause externalization even without any pinna cues (Hendrickx et al., 2017a; Plenge, 1972). Such dynamic binaural cues would be nearly identical in SHM or HRTF (dummy head or individual) rendering.

Even though the diotic-static as well as the anechoic stimuli received (relatively) high externalization ratings, the large standard deviations indicate strong differences in perceived externalization of these signals between the listeners.

In addition to several acoustic factors beneficial for high externalization ratings (visual and acoustical congruency, realistic presentation of reverberation, accurate simulation of binaural and spectral cues), the wording of the "At the position" rating of the externalization rating scale itself might have introduced a bias in conjunction with the listeners' tasks: Given that just before rating the externalization the listeners had indicted the perceived distance of the sound source, it could be considered consistent to respond that the perception was "At the position". While such a bias towards high externalization cannot be disproven, 8 of 10 listeners rated at least one stimulus per condition as perceived intern.

### *1. Extended statistical analysis*

Two listeners rated all stimuli in the reverberant conditions perceived as "At the position" (> 98 % and 100 %), and > 99 % and > 95 % as externalized, including anechoic conditions. To assess the effect of these listeners, the statistical analysis was repeated with these two listeners excluded. No change of the general statistical results was observed. Despite increasing the difference of rated externalization between diotic-static and the other rendering techniques, the standard deviation remained large (especially for diotic rendering). This still indicated a generally high externalization, with only a few listeners not externalizing the diotic-static rendered stimuli.

A second observation was that variances increased with higher externalization percentages. This suggest a logarithmic transformation of the data to obtain more equal variances. For the analysis of logarithmic-transformed results without the above two listeners, rendering became a significant factor in set A: [$F(3,21) = 3.157$, $p = 0.046$] and in set B: [$F(3,21) = 4.048$, $p = 0.020$]. However, pairwise comparisons revealed no significant differences. No other statistical results were changed by the transformation.

## C. Visual distance and room size estimation

Underestimations of visual distances (Kelly et al., 2022; Kelly, 2023; Korshunova-Fucci et al., 2023) were rarely observed in this study. The small room was overestimated by 8.75 % and the big room by 11.26 %. Only in the invisible room, the length of the purple line was underestimated (by 50.83 %). For the visual distance indication task (see Fig. 5), the underestimation was more pronounced. However, with an underestimation of -12.5 % for 5 m in the small room as well as an overestimation of +2.6 % for 5 m and an underestimation of - 5.4% for 10 m in the big room, the underestimation is still below the values reported in literature and even overestimation occurred. Even though the size of the invisible room was underestimated by roughly 50 %, for the indicated distances in that room, the 5 m distance was only overestimated by 10.2 %. Larger overestimations occurred for indicating the 10 m distance (+21 %) and 15 m distance (+54 %). These overestimations of the distance are likely caused by the underestimation of the room size itself. Further, the larger standard deviations in the invisible room (Sec. III. C.) indicate a stronger difference in perceived visual distance between the listeners compared to the visible rooms. Despite the visual differences between the here considered rooms, the 5 m distance could be estimated relatively accurately in all rooms. For the experiment conducted, this reflects the most important distance range, comparable to the simulated maximum distance of sound sources of 6 m.

## D. Distance judgement strategies

To ensure the validity of the visual aspects in this study, the listeners were asked to rate whether their eyes were open during the measurements. The high median rating of 1 and no rating lower than 2 demonstrates that this was the case, and the different visual conditions (LS, room visibility, partially room size) can be evaluated accordingly.

The dynamic head movement was used less often than anticipated. It was observed that dynamic motion was used more in the beginning of the experiment and its use declined during the experiment. A likely explanation is that the listeners got used to the stimuli and did not think that the movement would provide more helpful cues other than loudness and DRR, which are present in all stimuli even without motion.

While both, dynamic movement and the loudspeaker model were rated nearly identical in usefulness, only the visible loudspeaker statistically significantly improved the distance estimation. The fact that 11 out of 14 mentioned strategies are based on intuition or comparison of stimuli with each other, an imaginary loudspeaker (in conditions without LS) or the laser pointer on the floor suggests that listeners were not confident in identifying the position of the individual sources.

### E. Limitations and considerations for follow-up experiments

The bending of the Bézier curves fitted to the ADP results are a good indicator for the non-linearity of the distance perception with increasing distance, however, their usefulness in statistical analysis is limited at best. The analysis of slope would be sufficient, as not only did the analysis of bending give no additional results, it also failed to indicate significant effects as shown by the analysis of slope.

In follow-up experiments different white or pink noise burst instead of a frozen noise should be used to limit a somewhat unnatural direct comparison of identical source stimuli. Moreover, more natural, frequency-dependent reverberation (higher reverberation time at lower frequencies) would be more realistic for most environments and might improve the effectiveness of the DRR-cue by a more realistic spectral shape of the reverberation.

A more realistic (visual) room design (e.g., adding some furniture) could be used to help reduce potential differences in the visual perception of room dimensions between the listeners. In the current rooms with flat grey walls, listeners had to rely on stereoscopic, perspective, and parallax cues (when moving the head). Similarly, to promote size cues, the loudspeaker (used for LS model to indicate distances) could be presented to the listeners before the experiment, as to provide a real-life reference.

To better ensure persistent use of head motion during the experiment, it should be stated as mandatory during the instruction and either direct observation of the listeners or saving the head-tracked data for later analysis is advised.

## V.   CONCLUSIONS

Auditory distance perception (ADP) and externalization were evaluated in different virtual audio-visual environments for a frontal invisible sound source. The environments consisted of a big room with long reverberation time, a small room with short reverberation time, as well as a visually infinite black space (invisible room) in which the acoustics of both rooms was presented. Differently rendered stimuli, ranging from static diotic (monophonic) over static dummy head HRTF based binaural rendering to interactive binaural renderings using real-time head tracking and either an extended spherical head model or dummy head HRTFs were presented with a simulated sound source at distances between 1 m and 6 m (0.5 m step size) from the listener. Additionally, the effect of a freely movable loudspeaker model to indicate the perceived distance was tested. The following conclusions can be drawn:

- The observed pronounced overestimation of distances for ADP, depending on the visual environment, suggests the ability of relative distant judgements, while absolute ADP is clearly modulated by the visual reference frame. The larger the visually perceived space, the farther distances are rated, despite identical acoustical presentations.

- The availability of a freely movable loudspeaker model to which the auditorily perceived distance could be compared, improved ADP.

- ADP was neither significantly affected by headphone rendering technique nor by dynamic head motion. This suggest a strong dominance of monaural cues such as level and direct-to-reverberant energy ratio known from the literature. Potential binaural cues including acoustic parallax as a consequence of limited lateral motion played no significant role.

- Externalization depended to a lesser degree on the visual room representation than ADP and longer reverberation increased the perceived externalization. Although clear differences in externalization were observed for the different rendering techniques, those differences were not significant. However, there was a clear trend that the diotic-static presentation resulted in more internalized perception, in contrast to ADP, which was not affected by presentation mode. In contrast to ADP, the freely movable loudspeaker model did not affect externalization.

- For headphone-based presentation of virtual acoustics in virtual audio-visual environments, distance perception and externalization are partly independent, suggesting that distance can be estimated and experimentally judged independent of perceived externalization.

**AKNOWLEDGEMENTS**

The authors thank Stefan Fichna and Henning Steffens for support with the experimental setup. This work was supported by the Deutsche Forschungsgemeinschaft, DFG SPP Audictive – Project-ID 444827755.

**AUTHOR DECLARATIONS**

**Conflict of Interest**

The authors have no conflicts to disclose.

**Ethics Approval**

All participants were financially compensated, participated voluntarily, and provided informed consent. The study was approved by the Ethics committee of the University of Oldenburg.

**DATE AVAILABILITY**

The data is available for the authors on request.

**APPENDIX A: HRIR resolution reduction**

An overview of the spatial resolution is shown in Table IV. The HRIRs were time-aligned using signal processing routines from AKtools (Brinkmann and Weinzierl, 2017) and the SUpDEq toolbox Version 5.0.0 (Arend & Pörschmann, 2022). The extracted time of arrivals were stored in the delay field of the SOFA file containing the time-aligned HRIRs.

TABLE IV. Ranges and resolutions of different HRTF sections.

| Section | Elevation range of section | Elevation resolution | Azimuth resolution |
|---|---|---|---|
| Front - Fine | -35°:0°:35° | 5° | 5° |
| Front - Medium | -40°: -60°, 40°:60° | 10° | 10° |
| Front - Rough | -75°: -90°, 75°:90° | 15° | 15° |
| Back - Fine | -20°:0°:20° | 10° | 10° |
| Back - Medium | -30: -45°, 30°:45° | 15° | 15° |
| Back - Rough | -60°: -90°, 60°: 90°, | 15° | 20° |

**APPENDIX B: Bézier curve fit**

A second order (quadratic) Bézier curve of the form

$$P(t) = (1-t)^2 P_1 + 2t(1-t)P_2 + t^2 P_3 \quad \text{(B1)}$$

Was fitted to the distance data of each listener, with $0 \leq t \leq 1$ and $P_1$ being the starting point, $P_2$ the control point and $P_3$ the end point of the curve. Three parameters intercept, slope and bending were used to define the points ($P_1, P_2, P_3$). The x-values of $P_1$ and $P_3$ were fixed at the presented source distances 1 and 6 m (represented by t = 0 and t = 1), the y-values were fitted by varying slope and intercept. The bending parameter changed the deviation of $P_2$ from the center point between $P_1$ and $P_3$ on a straight line, ranging between the y-value of $P_3$ (upper y range) at x = 1 m for a bending parameter of +1 and the y-value of $P_1$ (lower y-range, bending parameter of –1 at x = 6 m). For the bending parameter a bounded parameter optimization (range -1 to +1) was performed. The parameter optimization was performed by MATLAB´s build in "fminsearch" function. An example fit is shown as cyan line in Fig. 2h).